\documentclass[pra,aps,twocolumn,amsfonts,showpacs]{revtex4}

\usepackage{graphicx}
\usepackage{epsfig}
\def\QED{\leavevmode\unskip\penalty9999 \hbox{}\nobreak\hfill
     \quad\hbox{\leavevmode  \hbox to.77778em{%
               \hfil\vrule   \vbox to.675em%
               {\hrule width.6em\vfil\hrule}\vrule\hfil}}
     \par\vskip3pt}
\def\qed{\leavevmode\unskip\penalty9999 \hbox{}\nobreak\hfill
     \quad\hbox{\leavevmode  \hbox to.77778em{%
               \hfil\vrule   \vbox to.675em%
               {\hrule width.6em\vfil\hrule}\vrule\hfil}}
     \par\vskip3pt}

\def\ibb #1{\leavevmode\hbox{\kern.3em\vrule
     height 1.5ex depth -.1ex width .4pt\kern-.3em\rm#1}}

\newcommand{\tr}[1]{{\rm tr}\left[#1\right]}

\newcommand\beq{\begin{equation}}
\newcommand\eeq{\end{equation}}
\newcommand\bea{\begin{eqnarray}}
\newcommand\eea{\end{eqnarray}}

\newtheorem{lemma}{Lemma}

\def\braket #1#2{\langle #1 , #2\rangle}
\def\ketbra #1#2{\vert #1\rangle \! \langle #2\vert}
\def\proof{\par\noindent{\it Proof:\ }}

\begin{document}

\title{Efficient distillation beyond qubits}

\author{Karl Gerd H. Vollbrecht}
\email{k.vollbrecht@tu-bs.de}
\author{Michael M. Wolf}
\email{mm.wolf@tu-bs.de} \affiliation{Institute for Mathematical
Physics, TU Braunschweig, Germany}

\date{\today}

\begin{abstract}
We provide generalizations of known two-qubit entanglement
distillation protocols for arbitrary Hilbert space dimensions. The
protocols, which are analogues of the hashing and breeding
procedures, are adapted to bipartite quantum states which are
diagonal in a basis of maximally entangled states. We show that
the obtained rates are optimal, and thus equal to the distillable
entanglement, for a $(d-1)$ parameter family of rank deficient
states. Methods to improve the rates for other states are
discussed. In particular, for isotropic states it is shown that
the rate can be improved such that it approaches the relative
entropy of entanglement in the limit of large dimensions.
\end{abstract}


\maketitle

\section{Introduction}

The possibility of {\it entanglement distillation} plays a crucial
role in Quantum communication and Quantum information processing
(cf. \cite{Springer}). Together with quantum error correction it
enables all the fascinating applications provided by Quantum
information theory in the presence of a noisy and interacting
environment. It provides a method to overcome decoherence without
requiring a complete isolation from the environment.

The usual abstract way of thinking about entanglement distillation
is the following: Two parties, Alice and Bob, situated at distant
locations share $n$ copies of a mixed entangled quantum state
$\rho$, which they may have obtained by sending one part of a pure
maximally entangled state through a noisy quantum channel. Assume
that both parties are able to perform any collective quantum
operation, which merely acts locally on their part of the $n$
copies. Moreover, Alice and Bob are connected via a classical
channel, such that they can perform arbitrary many rounds of local
quantum operations, where each round may depend on the measurement
outcomes of all the preceding operations on both sides. The set of
operations accessible in this way is called LOCC (local operations
and classical communication).

It was shown in one of the seminal works from the early years of
quantum information theory \cite{BennettPRL} that under the above
conditions Alice and Bob can for certain two-qubit states {\it
distill} a smaller number $m$ of states $\rho'$, which are closer
to the maximally entangled state than the initial ones, from a
larger number $n\geq m$ of weaker entangled states $\rho$. This
can be done in such a way that in the limit $n\rightarrow\infty$
the output states $\rho'$ become maximally entangled. The
asymptotic ratio $m/n$ is then called the {\it rate} of the
distillation protocol and the maximally accessible rate under all
LOCC protocols is an important measure of entanglement, the {\it
distillable entanglement} $D(\rho)$. The latter quantifies in some
sense the amount of useful entanglement contained in the state
$\rho$.

Up to now the {\it hashing/breeding} distillation protocol
presented in \cite{BennettPRL,BennettPRA}, which is adapted to
Bell diagonal states of two qubits, is essentially the only
protocol leading to a non-zero rate in the asymptotic limit.

The present paper is devoted to generalize these protocols to
higher dimensions. We will thereby closely follow the ideas of
Ref.\cite{BennettPRL,BennettPRA}. The protocols are adapted to
states which are diagonal in a basis of maximally entangled states
or mapped onto such states by an LOCC {\it twirl} operation. The
main step is to translate the apparent quantum task into a
classical problem in such a way that all the operations involved
are LOCC. The following results are obtained:
\begin{itemize}
  \item We provide an entanglement assisted distillation protocol
  ({\it breeding}) that works for any finite dimension $d$. It is in this case assumed
  that Alice and Bob share maximally entangled states initially, which they may
  have obtained with a different protocol (e.g. hashing in a prime dimensional subspace)
  and with a smaller rate. The attained
  distillation rate is given by
\begin{equation}\label{rate}
  \log_2 d - S\big({\cal T}(\rho)\big),
\end{equation}where $S\big({\cal T}(\rho)\big)$ is the von Neumann entropy of the twirled state $\rho$.
  \item We show that a {\it hashing protocol} exists for prime  dimensions $d$, which does
  not require any predistilled states. The obtained rate is the same as in Eq.
  (\ref{rate}). The method can easily be generalized to dimensions
  which are powers of primes, if one uses a different basis of
  maximally entangled states.
  \item Both protocols, breeding and hashing, are shown to be
  optimal for a $(d-1)$ parameter family of low rank states. This
  generalizes the observation of Rains \cite{REbound2} for the case $d=2$.
  \item We discuss methods for improving the rates for Isotropic
  states. In particular the projection onto local subspaces will
  turn out to yield the optimal rate in the limit of large
  dimensions.
\end{itemize}

Despite the practical relevance and quite considerable effort in
the theory of entanglement distillation, many of the basic
questions are yet unanswered. So do we neither know a decidable
necessary and sufficient criterion for distillability in Hilbert
spaces $\mathbb{C}^d\otimes\mathbb{C}^d$ with $d>2$, nor do we
know $D(\rho)$ even for otherwise simple quantum states $\rho$.
What makes the investigation of distillability so difficult is on
the one hand the asymptotic limit ($n\rightarrow\infty$) and on
the other hand the mathematically intractable set of LOCC
operations.

However, many partial results have been obtained in various
directions, and before we go into the details of the present
article we want at least to briefly recall some of them:

\paragraph{Distillability: }

For the case of two qubits it was shown in \cite{2Q} that every
entangled state is distillable. A general necessary condition for
the distillability
   of a state  $\rho$ is the fact that its partial transpose
$\rho^{T_A}$, defined with respect to a given product basis by
$\langle ij|\rho^{T_A}|kl\rangle=\langle kj|\rho| il\rangle$, has
a negative eigenvalue \cite{HHH2}. Except for special cases like
states on ${\mathbb{C}}^2\otimes{\mathbb{C}}^n$ \cite{2n,IHa} and
Gaussian states \cite{Gauss} it is however unclear whether this
condition is sufficient as well. There is some evidence presented
in \cite{IHa,IBM} that this may not be the case and that there are
indeed undistillable states, whose partial transpose is not
positive (NPPT). However, it was shown in \cite{VWPRL} (see also
\cite{otheractivations}) that every NPPT state becomes distillable
when adding a certain {\it bound entangled} state with positive
partial transpose (PPT). Moreover, this {\it activation} process
was shown to require only an infinitesimal amount of entanglement
contained in the additional PPT state \cite{VWPRL}.

\paragraph{ Distillation protocols: }
Up to now, essentially the only mixed state distillation protocol
leading to a non-zero rate in the asymptotic limit is the
breeding/hashing protocol presented in
\cite{BennettPRL,BennettPRA}, which is adapted to Bell-diagonal
states of two qubits. There are, however, several purification
schemes which increase the purity and entanglement of a state
without yielding a non-zero rate on their own. The best known
examples are the {\it recurrence} protocols presented in
\cite{BennettPRL,BennettPRA} (see \cite{imperfect} for the
discussion of imperfect operations) and \cite{Ekert}. These were
investigated also for higher dimensions in \cite{HHHred} and
\cite{AlberJex}. In \cite{Smolin} it was shown that these methods
can be improved by using more than one source state per target. A
systematic generalization of the local unitaries applied in the
recurrence method for two qubit systems has recently been provided
in \cite{Frank}. A protocol which is analog to the hashing
procedure but uses PPT preserving rather than LOCC operations is
described in \cite{Rainssemi}.

\paragraph{ Distillable Entanglement: }
For pure states the distillable entanglement is equal to the von
Neumann entropy of the reduced state \cite{Bennettpure}. For mixed
states only bounds are known, which can be calculated for some
very special cases (cf. \cite{OO}). The best known upper bound for
the distillable entanglement can be found in \cite{Rainsbound}. A
closely related bound, which we will describe in detail later, is
given by the {\it relative entropy of entanglement}
\cite{REbound1,REbound2}.

\paragraph{ Relations to other Quantum information tasks: } It was
already noticed in \cite{BennettPRA} that a distillation protocol
involving only one-way classical communication (like the hashing
protocol) corresponds to a quantum error correcting code and vice
versa. Moreover, upper bounds for the distillable entanglement
allow one to obtain upper bounds for quantum channel capacities
(cf. \cite{HHHchannelcap}). In particular, the inequality
\begin{equation}\label{Hashineq}
D_1(\rho)\geq S\big({\rm tr}_B[\rho]\big)-S(\rho),
\end{equation}
where $D_1(\rho)$ is the distillable entanglement for one-way
communication protocols, implies Shannon-like formulas for quantum
capacities, providing the quantum noisy coding theorem
\cite{HHHchannelcap}. The inequality (\ref{Hashineq}) is
consistent with the results we obtain in Sec. \ref{lowrank}.

An important application of distillation and purification schemes
is also the possibility of factoring out an eavesdropper for
secret quantum communication \cite {Ekert,BriegelPAA}. This is
known as {\it Quantum privacy amplification}. \hspace*{10pt}

\section{Preliminaries}\label{sec1}

Let us first introduce the preliminaries we need for the
distillation protocols described in Sec. \ref{breed} and
\ref{hash}. We begin with characterizing the states for which the
protocols are adapted and we will subsequently specify the set of
required local operations.

\subsection{Basis of maximally entangled states}
Bases of maximally entangled states $\{\Psi_{kl}\}$,
$k,l\in\{0,\ldots,d-1\}$ exist for any Hilbert space ${\cal
H}=\mathbb{C}^d\otimes\mathbb{C}^d$ and correspond to an
orthonormal operator basis of $d^2$ unitary $d\times d$ matrices
$\{U_{kl}\}$ via
\begin{equation}\label{UPsi}
|\Psi_{kl}\rangle \equiv \left({\bf 1}\otimes U_{kl}\right)
|\Omega\rangle,
\end{equation}
with $|\Omega\rangle$ being a maximally entangled state. The
latter is in the following chosen to be $|\Omega\rangle =
|\Psi_{00}\rangle = \frac1{\sqrt{d}}\sum_{j=0}^{d-1} |jj\rangle$.
A general construction procedure \cite{teledens} for unitary
operator bases involves Latin squares and complex Hadamard
matrices and the best known example constructed in this way is of
the form
\begin{equation}\label{Us}
U_{kl}= \sum_{r=0}^{d-1} \eta^{rl} |k\oplus r\rangle\langle r|\
,\quad \eta=e^{\frac{2\pi i}d}
\end{equation}
where $\oplus$ means addition modulo $d$. The set of unitaries in
(\ref{Us}) is orthonormal with respect to the Hilbert-Schmidt
scalar product, i.e., $\tr{U_{ij}^*U_{kl}}=d
\delta_{ik}\delta_{jl}$, and forms a discrete Weyl system since
$U_{ij}U_{kl}=\eta^{jk} U_{i\oplus k,j\oplus l}$.

In the following we will solely use the maximally entangled basis
from Eqs.(\ref{UPsi},\ref{Us}) or tensor products thereof and
denote the first index as the {\it shift} and the second as the
{\it phase} index.

\subsection{Symmetric states}
Symmetric states commuting with a group of local unitaries play an
important and paradigmatic role in Quantum information theory and
in particular in the context of entanglement distillation. The
best known examples are {\it Werner states}, {\it Isotropic
states} and {\it Bell diagonal states}. The latter are convex
combinations of the four maximally entangled Bell states of two
qubits and contain in this case the Isotropic states, which are
for two qubit systems in turn equivalent to Werner states.

In analogy to the two qubit case, we will call a state  on
$\mathbb{C}^d\otimes\mathbb{C}^d$ {\it Bell diagonal } if it can
be written as a convex combination of maximally entangled states
$P_{ij}=|\Psi_{ij}\rangle\langle\Psi_{ij}|$. The corresponding
symmetry group is given by
\begin{equation}\label{G}
G=\{U_{i,j}\otimes U_{i,-j}\},
\end{equation}
which is an abelian group with the property that its commutant,
which is again spanned by $G$, contains $d^2$ one-dimensional
projectors $\{P_{ij}\}$. Hence, for $\{\lambda_{ij}\}$ being
convex weights:
\begin{equation}\label{Geqi}
\rho=\sum_{ij}\lambda_{ij} P_{ij}\ \Leftrightarrow \ \forall g\in
G: [\rho,g]=0.
\end{equation}
Moreover, every not symmetric state $\rho$ can be mapped onto a
Bell diagonal state ${\cal T}(\rho)$ by means of a discrete twirl
operation
\begin{equation}\label{dtwirl}
{\cal T}(\rho)=\frac1{d^2}\sum_{g\in G} g^* \rho g,
\end{equation}
which can be implemented by means of local operations and
classical communication (LOCC).

Isotropic states, which are completely characterized by their {\it
fidelity} $f:=\langle\Omega|\rho|\Omega\rangle$, are a special
instance of Bell diagonal states with $\lambda_{00}=f$ and
$\lambda_{ij}=(1-f)/(d^2-1)$ for $(i,j)\neq (0,0)$. An Isotropic
state is entangled and distillable iff $f>\frac1d$ \cite{HHHred}.

\subsection{Local operations and measurements}
An essential ingredient for any distillation protocol acting on
two qubit systems is the CNOT operation. One possibility of
generalizing this operation to higher dimensions is to use a {\it
controlled shift} operation (CS) \cite{HHHred}, which acts as
\begin{equation}\label{Cshift}
C |i\rangle\otimes|j\rangle = |i\rangle\otimes |j\oplus i\rangle,
\end{equation}
where the first tensor factor is the {\it source} and the second
the {\it target}. It is readily verified that a {\it bilateral}
controlled shift operation (BCS), where both parties in a bipartite
system apply CS operations locally on a tensor product of two
maximally entangled states, acts as
\begin{equation}\label{BCS}
(C\otimes C ) |\Psi_{ij}\rangle\otimes |\Psi_{kl}\rangle =
|\Psi_{i,j\ominus l}\rangle\otimes|\Psi_{k\oplus i,l}\rangle,
\end{equation}
where the first tensor product on the l.h.s. in (\ref{BCS})
corresponds to the Alice$|$Bob split, whereas the others
correspond to the source$|$target split. Note that the target
state picks up the shift index of the source, i.e., $k\mapsto
k\oplus i$, while the source remains unchanged iff $l=0$.

If we measure a target state of the form $|\Psi_{kl}\rangle$ in
computational basis, then $k$ is the difference of the measurement
outcomes. In this way we can obtain information about the shift
index of an unknown source state, if we know the target state: We
first apply a BCS operation and then measure the target state in
computational basis.

An analogous transformation can be defined for the phase indices,
since we can simply interchange the two indices by the following
local unitary operation:
\begin{equation}\label{Had}
\big(V \otimes \overline{V}\big)P_{ij}\big(V \otimes
\overline{V}\big)^* = P_{j,-i} ,\quad V=\frac1{\sqrt{d}}\sum_{k,l}
\eta^{kl}|k\rangle\langle l|.
\end{equation}
So we can define a {\it modified bilateral controlled shift
operation} (mBCS) by first applying a $V \otimes \overline{V}$
transformation to the source state, followed by a  BSC operation
on source and target and then undoing the local operation on the
source state by a $(V \otimes \overline{V})^*$ rotation. The mBCS
operation then acts as
\begin{equation}\label{mBCS}
P_{ij}\otimes P_{kl} \mapsto P_{i\oplus l,j}\otimes P_{k\oplus
j,l},
\end{equation}
where the second tensor factor corresponds to the target again.

Assume now we have one target state $|\Psi_{00}\rangle$ and $n$
source states with shift indices $k_1,\ldots,k_n$  and phase
indices $l_1,\ldots,l_n$. We now apply $s_a$ BCS operations and
$p_a$ mBCS operations to the $a$-th source and the target. If we
then measure the target state again in computational basis, the
difference of the measurement outcomes will be
\begin{equation}\label{diffshift}
\Big(\bigoplus_{a=1}^n k_a s_a\Big) \oplus \Big(\bigoplus_{a=1}^n
l_a p_a\Big)  =:\langle \vec{k},\vec{s}\rangle \oplus\langle
\vec{l},\vec{p}\rangle.
\end{equation}
This enables us to obtain some information about the distribution
of the shift and phase indices of the source states without
disturbing them.

In the following we will organize the shift and phase indices of a
sequence of maximally entangled states in a single vector $\vec
S:=(k_1,\ldots,k_n,l_1,\ldots l_n)$ and accordingly the
multiplicities of the BCS and mBCS operations in a vector $\vec
M:=(s_1,\ldots,s_n,p_1,\ldots p_n)$. Both $\vec S$ and $\vec M$
then belong to  $\{0,\ldots,d-1\}^{2n}$, i.e. they are elements of
the group $\mathbb{Z}_d^{2n}$ with addition modulo $d$.

The sequence of local operations characterized by $\vec M$
followed by a measurement of an additionally required target state
$P_{00}$ thus yields the information \beq\label{MSscalar}
\langle\vec M,\vec S\rangle :=\bigoplus_{i=1}^{2n} M_i S_i
=\langle \vec{k},\vec{s}\rangle \oplus\langle
\vec{l},\vec{p}\rangle \eeq about a sequence $\vec S$ of maximally
entangled states, without disturbing the latter.

To complete the list of later on required basic local operations
we have still to introduce a generalization of a $\pi/2$ rotation
given by \beq\label{ugdef} u(g):=\sum_{k} e^{\frac{i\pi}{d} k^2 g}
\ketbra{k}{k}, \eeq where $g\in\mathbb{Z}$ is an arbitrary number.
We will use this within bilateral operations of the form
$u(g)\otimes \bar u(g)$ and $v(g)\otimes \bar v(g)$, where
$v(g):=V^*u(g)V$. These unitary operations act on a maximally
entangled state $P_{kl}$ as \bea
\big[u(g) \otimes \bar u(g)\big] P_{kl} \big[u(g) \otimes \bar u(g)\big]^*=P_{k,l\ominus gk}, \\
 \big[v(g) \otimes \bar v(g)\big] P_{kl} \big[v(g) \otimes \bar v(g)\big]^*=P_{k\oplus gl,l}.
\eea

\subsection{Why primes are special}

Measurements based on the $\mathbb{Z}_d^{2n}$ scalar product in
Eq.(\ref{MSscalar}) will play a crucial role in the entanglement
distillation protocols discussed in the sequel. One of the main
tasks will thereby be to choose the {\it measurement vector} $\vec
M$ such that we obtain as much information about the {\it
sequence} $\vec S$ as possible. In general this can be a highly
nontrivial problem. However, if the dimension $d$ is a prime, then
choosing $\vec M$ randomly turns out to be a pretty good choice.
The reason why this works well only for prime dimensions is that
in this case $\mathbb{Z}_d$ is a {\it field}, which means that it
is not only an abelian group with respect to the addition modulo
$d$, but also with respect to the multiplication if we exclude the
zero element. That is, for $a,b,x \in \mathbb{Z}_d,\ a\neq 0$
every equation of the form
\begin{equation}\label{axb}
b\ =\ a\cdot x\ \text{ mod } d
\end{equation}
has a unique solution for $x$ if and only if $d$ is prime. This
leads us to the following lemma:

\begin{lemma}
Let $\vec S \neq \vec S'$ be elements of $\mathbb{Z}_d^{m}$ with
$d$ prime. Given a scalar product $\langle \vec x,\vec y\rangle :=
\bigoplus_{i=1}^m x_i y_i$ and a uniformly distributed random
vector $\vec M\in \mathbb{Z}_d^{m}$, the probability for $\langle
\vec M,\vec S\rangle = \langle \vec M,\vec S'\rangle$ is equal to
$1/d$.
\end{lemma}
\proof We ask for the probability that $\braket{\vec S-\vec
S'}{\vec M}=0$. Since $\vec S \neq \vec S'$, there exists a
component $x$ where $S_x\neq S_x'$ and we can write
\beq\label{SSxx} 0\equiv\braket{\vec S-\vec S'}{\vec
M}=M_x(S_x-S_x')+\sum_{i\neq x}M_i (S_i-S_i'). \eeq Assume now
that all $M_i, i \neq x$ are already randomly chosen. Then
Eq.(\ref{SSxx}) has the form of Eq.(\ref{axb}) with $x=M_x$,
$a=(S_x-S_x')$ and $b=\sum_{i\neq x}M_i (S_i'-S_i)$.  Since this
equation has a unique solution for $x$ and $M_x$ is a uniformly
distributed random variable, the probability that $M_x$ matches
the solution is indeed $\frac1d$. \QED

\section{The breeding protocol}\label{breed}
The {\it breeding} protocol is the preliminary stage of the {\it
hashing} protocol discussed in the next section. Both are adapted
to the distillation of Bell diagonal states and the main idea is
to transform this quantum problem into the classical problem of
identifying a word given the probability distribution of the
respective alphabet and a restricted set of measurements.

Assume that Alice and Bob share $n$ copies of a Bell diagonal
state $\rho$, such that
 \beq\label{thestate}
\rho^{\otimes n}=\sum_{k_1 \ldots k_n, l_1 \ldots l_n}
\lambda_{k_1l_1}\cdots\lambda_{k_n l_n} P_{k_1 l_1}\otimes
\cdots\otimes P_{k_n l_n}. \eeq An appropriate interpretation of
Eq.(\ref{thestate}) is to say that Alice and Bob share the
 state \beq P_{k_1 l_1}\otimes \cdots\otimes
P_{k_n l_n}:=P_{\vec S} \eeq with probability
$\lambda_{k_1l_1}\cdots\lambda_{k_n l_n}$. Note that if they knew
the sequence $\vec S=(k_1,\ldots,k_n,l_1,\ldots,l_n)$, they could
apply appropriate local unitary operations in order to obtain the
standard maximally entangled state $P_{00}^{\otimes n}$ and thus
gain $n \log_2 d$ ebits of entanglement.

Let us further assume that they have already a sufficiently large
set of predistilled maximally entangled states $P_{00}$ and that
they are able to perform all the local operations such as BCS and
mBCS operations described in the previous section. In this case
they can utilize the predistilled states for performing local
measurements leading to the result $\braket{\vec M}{\vec S}$ as
discussed in Eq.(\ref{MSscalar}) in order to obtain some
information about the sequence $\vec S$. However, every single
measurement $\vec M$ will destroy one of the predistilled
maximally entangled states, and the task is therefore to identify
the sequence $\vec S$ using as few measurements as possible.

At this point the quantum task of distillation is  transformed
into an entirely classical problem: Given an unknown vector $\vec
S$ and the possibility of measuring functions of the form
$\braket{\vec M}{\vec S}$, how many measurements $r$ do we need to
identify $\vec S$? Knowing $\vec S$ we gain $n$ maximally
entangled pairs, but destroyed $r$, such that the overall gain
would be $(1-r/n)\log_2 d$ ebits per copy.

We would need $r=2 n$ measurements to  identify a completely
random ${\vec S}\in \mathbb{Z}_d^{2n}$ with independently and
uniformly distributed components. So we would destroy more
entangled pairs than we gain.

Fortunately, however, the set of possible vectors $\vec S$ is not
completely random but has a distribution depending on the spectrum
of the Bell diagonal state $\rho$. In the limit of many copies
there exists thus a set of {\it likely sequences} containing only
$2^{n S(\rho)}\leq d^{2n}$ different vectors $\vec S$ with the
property that the probability that a vector $\vec S$ is contained
within this set of likely sequences approaches one in the limit
$n\rightarrow\infty$ \cite{likelysequences}. Here $S(\rho)$ is the
von Neumann entropy of the density matrix, which is equal to the
Shannon entropy of the spectrum of $\rho$, i.e.
\beq\label{Srho}S(\rho)=-\sum_{k,l} \lambda_{k l}\log_2\lambda_{k
l}.\eeq

Now let Alice and Bob choose each measurement $\vec M$ randomly.
Then Lemma 1 tells us that if $d$ is prime, they can reduce their
list of possible sequences by a factor of $\frac1d$ after every
measurement by deleting every sequence, which is not consistent
with the obtained result. Since their list initially contains
$2^{n S(\rho)}$ possible sequences, $r=n S(\rho)/\log_2d$
measurements are required in order to identify $\vec S$ in the
limit of large $n$. The rate obtained in this way in the
asymptotic limit is a lower bound to the (entanglement assisted)
distillable entanglement which is thus \beq\label{erg} D(\rho)\geq
\log_2 d - S(\rho).
 \eeq

Let us now extend the obtained result, expressed in
Eq.(\ref{erg}), from prime dimensions to arbitrary ones. As we
have already mentioned, Lemma 1 holds only for primes and cannot
be extended. A way out of this dilemma is to use controlled shift
operations between Hilbert spaces of different dimensions. The
definition looks similar to Eq.(\ref{Cshift2}):
\begin{equation}\label{Cshift2}
C' |i\rangle\otimes|j\rangle := |i\rangle\otimes |j\oplus
i\rangle,
\end{equation}
with the only difference that $i$ runs from $0$ to $d-1$ whereas
$j$ runs from $0$ to $d'-1$, where $d'$ is now supposed to be a
prime with $d'\geq d$. Moreover, the addition in the second ket
(target) is modulo $d'$.

The penalty of the BCS and mBCS operations based on such a
controlled shift operation is, that the source states after the
operation will in general no longer be of the form $P_{kl}$ given
by Eqs.(\ref{UPsi},\ref{Us}). However, they still are, if the
target state, acting on $\mathbb{C}^{d'}\otimes\mathbb{C}^{d'}$,
has a phase index equal to zero. Then:
\begin{equation}\label{Bshift2}
(C'\otimes C') |\Psi_{ij}\rangle\otimes |\Psi_{k 0}'\rangle =
|\Psi_{i,j}\rangle\otimes|\Psi_{k\oplus i,0}'\rangle.
\end{equation}
In this way we can straightforward generalize the result of
Eq.(\ref{erg}) to Bell diagonal states of arbitrary dimension,
just by reshuffling the degrees of freedom of the predistilled
maximally entangled states. Hence, the crucial point is, that the
target states, which are in the case of the breeding protocol the
predistilled resources, have prime dimensions.

\section{The hashing protocol}\label{hash}
 So far we have assumed that Alice
and Bob initially share a set of predistilled maximally entangled
states. We took into account this resource when calculating the
rate, but since the additivity of the distillable entanglement is
not yet decided for bipartite systems, we might expect a smaller
rate for the (non entanglement assisted) distillation rate. The
{\it hashing protocol}, however, achieves the same asymptotic rate
as the previously discussed breeding protocol without requiring
additional predistilled states. The main idea is though still the
same: Alice and Bob identify the sequence $\vec S$ by means of
LOCC operations.

The protocol, however, is now a bit more complicated since if we
use one of the $n$ states characterized by $\vec
S\in\mathbb{Z}_d^{2n}$ as target for the BCS and mBCS operations
then we have to take into account the {\it backaction} from the
unknown target to the remaining source states.

We will in the following choose the $n$-th state characterized by
$S_n$ and $S_{2n}$ as the target and consider the remaining $n-1$
states characterized by a vector $\vec
R=(S_1,\ldots,S_{n-1},S_{n+1},\ldots,S_{2n-1})$ as source states.

In the first step of the protocol a bilateral $ v(g)\otimes \bar
v(g)$ rotation with a randomly chosen number $g\in \{0, \ldots,
d-1\}$ is applied to the target state. This shifts some
information about the phase index of the target state into its
shift index.

Then a random vector $\vec M\in\mathbb{Z}_d^{2n-2}$ is chosen and
the two parties perform the respective BCS and mBCS operations.
After measuring the target state, the difference of the
measurement outcomes, i.e. the final shift index of the target,
will be \beq\label{out2}\langle\vec M;g | \vec S\rangle:=
\braket{\vec M}{\vec R}-\sum_{i =1}^{n-1} M_i M_{i+n-1}
S_{2n}+S_n+gS_{2n}. \eeq The first term $\braket{\vec M}{\vec R}$
in Eq.(\ref{out2}) is the familiar outcome if the target state is
$P_{00}$. The terms $-\sum_{i =1}^{n-1} M_i M_{i+n-1} S_{2n}+S_n$
correspond to the other possible targets and their backaction onto
the source states, and the last term $gS_{2n}$ is due to the
initial $v(g)\otimes \bar v(g)$ rotation.

  Note that in general the
outcome also depends on the order in which the different BCS and
mBCS operations are carried out. For Eq.(\ref{out2}) first all BCS
and then all mBCS operations were applied.

With the possibility of measuring functions of the form
(\ref{out2}) we will now follow the line of argumentation of the
previous section and show that the gain of information  is
(asymptotically) at least $\log_2 d$ bits per measurement, if the
dimension $d$ is prime. Therefore we ask again for the probability
that two different sequences $\vec S\neq\vec S'$ lead to the same
measurement outcome under the assumption that $\vec M$ and $g$ are
uniformly distributed random variables. To this end we have to
distinguish between three different cases:
\begin{enumerate}
  \item The two sequences only differ in $S_n\neq S_n'$: This is
  the trivial case where $\vec S$ and $\vec S'$ are distinguished
  with unit probability.
  \item $\vec S$ and $\vec S'$ differ in the phase index of their target, i.e. $S_{2n}\neq
  S_{2n}'$: If we assume in this case that $\vec M$ is already chosen, then the equation
  $\langle \vec M;g|\vec S\rangle=\langle \vec M;g|\vec S'\rangle$ is of the form $b=a\cdot
  x$ with $g$ playing the role of $x$. Hence, the probability that
  a randomly chosen $g$ matches the right solution is $\frac1d$, if $d$ is prime.
  \item $\vec S$ and $\vec S'$ have a target state with the same phase index, i.e.
  $S_{2n}=S_{2n}'$ and $\vec R\neq\vec R'$: Then Lemma 1 tells us, that a
  randomly chosen $\vec M$ distinguishes the two sequences $\vec R\neq\vec
  R'$ with probability $\frac1d$, if $d$ is prime.
\end{enumerate}

The hashing protocol consists now of several rounds of such
measurements, where each round destroys the entanglement of one of
the $n$ pairs characterized by $\vec S$. The relevant part of the
system after $r$ rounds is thus described by a vector $\vec
S(r)\in\mathbb{Z}_d^{2(n-r)}$ and the measurements applied to
$\vec S(r)$ are in turn characterized by $\vec
M(r)\in\mathbb{Z}_d^{2(n-r-2)}$ and $g(r)\in \mathbb{Z}_d$. Note
that given the sequence of measurements, the vector $\vec S(r)$
deterministically depends on $\vec S=\vec S(0)$.

The above discussion implies now, that the probability
\begin{eqnarray}
&&P\Big[\vec S(r)\neq\vec
S(r)'\wedge\nonumber\\
&&\quad\forall_{k=0}^{r-1}: \langle\vec M(r);g(r)|\vec S(r)-\vec
S(r)'\rangle=0\Big]\label{PSS'}
\end{eqnarray}
that $\vec S(r)$ and $\vec S(r)'$ have agreed on all $r$
measurement results and thereby remain distinct, is at most
$d^{-r}$, if $d$ is prime. Since we have initially again about
$2^{n S(\rho)}$ likely sequences, we need (in the limit of large
$n$) $r=n S(\rho)/\log_2 d$ rounds in order to identify the
remaining sequence $\vec S(r)$. This leads us again to the rate in
Eq.(\ref{erg}).

What if $d$ is no prime? Unfortunately, we cannot extend the
result to other dimensions in the way we did in the case of the
breeding protocol, since we do not have any target states of prime
dimension with zero phase indices. However, if the dimension is a
power of a prime $d=d'^p$, then we can easily extend the result
for states which are diagonal with respect to a different basis of
maximally entangled states, given by:
\begin{equation}\label{Pveckl}
P_{\vec k \vec l}:=\bigotimes_{i=1}^p P_{k_i l_i},
\end{equation}
where $P_{k_i l_i}$ is a maximally entangled state acting on
$\mathbb{C}^{d'}\otimes\mathbb{C}^{d'}$. The (LOCC) twirl
operation, which maps an arbitrary state onto a state diagonal in
this basis is given by $\rho\mapsto{\cal T}^{\otimes p}(\rho)$.

The hashing protocol can then be applied in the described manner
to $p n$ tensor factors of prime dimension $d'$ and we yield in
this way again the rate in Eq.(\ref{erg}) for states in $d=d'^p$
dimensions. Moreover, note that isotropic states are diagonal in
both bases, given by Eqs.(\ref{UPsi},\ref{Us}) and
Eq.(\ref{Pveckl}).

\section{Optimality for low rank states}\label{lowrank}

It is well known from the case of two qubit systems that the
hashing/breeding protocol is not optimal in general. However, for
a certain $(d-1)$ parameter family of rank deficient states, we
will show that the obtained rate is equal to the relative entropy
of entanglement. Since the latter is an upper bound to the
distillable entanglement, this implies that the protocols are
optimal in this case. For $d=2$ this was first noticed by Rains in
\cite{REbound2}.

Consider mixed states of rank smaller than or equal $d$
\begin{equation}\label{rhomu}
  \rho_{\mu} = \sum_{l=1}^d \mu_l |\phi_l\rangle\langle \phi_l|,
\end{equation} which are diagonal with respect to
\begin{equation}\label{phil}
|\phi_l\rangle := |\Psi_{0,l}\rangle =
\frac1{\sqrt{d}}\sum_{r=0}^{d-1} e^{\frac{2\pi i}d rl} |r
r\rangle.
\end{equation}
The set of these states forms a simplex in $\mathbb{R}^{d-1}$ and
has the property that every state $\rho_{\mu}$ except for the
barycenter, for which $\mu_l=\frac1d$, is entangled. To see this,
note first that $\max_l\{\mu_l\}$ is the fully entangled fraction.
That is, if not all $\mu_l$ are equal, then the fully entangled
fraction is larger than $\frac1d$ and the respective state is thus
entangled. On the other hand, if $\mu_l=\frac1d$, then
\begin{eqnarray}
\rho_\mu&=&\frac1{d} \sum_{l=1}^d |\phi_{l}\rangle\langle\phi_{l}|\label{rhosep}\\
&=&\frac{1}{d^2}\sum_{l,r,s} \exp\Big[\frac{2\pi i}{d} l(r-s)
\Big] |r, r\rangle\langle s, s|\\ &=& \frac1d \sum_{r=0}^{d-1}
|r\rangle\langle r|\otimes |r\rangle\langle r|
\end{eqnarray} is evidently a separable state, which is said to be {\it maximally correlated}.
We will denote this state by $\rho_{\text{sep}}$ in the sequel.

An upper bound to the distillable entanglement is given by the
{\it relative entropy of entanglement} \cite{REbound1,REbound2},
which is defined by
\begin{equation}\label{RelEntdef}
E_R(\rho)=\inf_{\sigma}
\Big(\tr{\rho(\log_2\rho-\log_2\sigma)}\Big),
\end{equation}where the infimum is taken over all separable states
$\sigma$. Hence, if we choose $\sigma=\rho_{\text{sep}}$, the
distillable entanglement is bounded by
\begin{equation}\label{Ineqchain}
D(\rho_{\mu})\leq E_R(\rho_{\mu})\leq
-\tr{\rho\log_2\rho_{\text{sep}}} - S(\rho_{\mu}).
\end{equation}
However, $-\tr{\rho\log_2\rho_{\text{sep}}}=\log_2d$ and the rate
achieved by the breeding/hashing protocol is thus optimal for
every $\rho_{\mu}$. Strictly speaking, the hashing protocol leads
to the distillable entanglement and the breeding rate is equal to
the  distillable entanglement $D'$ assisted by a maximally
entangled resource $\omega$. The latter can easily be seen by
noting that
\begin{eqnarray}\label{Dprimebound}
D'(\rho)&=&D(\rho\otimes\omega)-E_R(\omega)\\ &\leq&
E_R(\rho\otimes\omega)-E_R(\omega)\leq E_R(\rho).
\end{eqnarray}
That is, the relative entropy of entanglement is an upper bound
for the entanglement assisted distillable entanglement as well.

\section{Improving the rates for Isotropic states}
For a general Bell diagonal state the introduced protocols are not
optimal. In particular for states with a large entropy the rate is
poor or even zero. In order to improve this, one can make use of
hybrid protocols, where the first step decreases the entropy while
 conserving most of the entanglement, and in a second step the
 hashing/breeding protocol is applied. There are many ways of
 performing such an entropy decreasing preprocessing. We will in
 the following discuss two of them for the case of Isotropic
 states: The {\it recurrence method} which is well known for qubits \cite{BennettPRL,BennettPRA,Ekert} and has already been
 investigated by \cite{HHHred,AlberJex} for $d>2$, and the {\it projection onto local
 subspaces}. For the latter we will show that in the limit of
 large dimensions the rate approaches the relative entropy of
 entanglement. Hence, the achieved rate is optimal in that limit.

{\it Isotropic states}, for which we will discuss both methods in
more detail, have the form
\begin{equation}\label{Isotropic}
\rho=f |\Omega\rangle\langle\Omega|+\frac{1-f}{d^2-1}\big({\bf
1}-|\Omega\rangle\langle\Omega| \big).
\end{equation}
That is they are depolarized maximally entangled states depending
on a single {\it fidelity} parameter $f\in [0,1]$. The twirl
operation mapping every state onto an isotropic state has the form
$\rho\mapsto \int dU (U\otimes\overline U)\rho (U\otimes\overline
U)^*$, i.e. it is an averaging over the unitary group $U(d)$ with
respect to the Haar measure $dU$. The states in
Eq.(\ref{Isotropic}) are entangled (and distillable) iff
$f>\frac1d$ \cite{HHHred}.

\subsection{The recurrence method}
The recurrence method for higher dimensions and variations thereof
have already been discussed in detail in \cite{HHHred} and
\cite{AlberJex}. For completeness, however, we will recall the
main steps. The idea is to apply a BCS operation on two copies of
the state and then to measure the target state in computational
basis. The source states are kept whenever the measurement
outcomes coincide, otherwise they are discarded. The remaining
states may then be twirled onto isotropic states again and we can
proceed either with iterating the recurrence method or applying
the hashing/breeding procedure if the entropy of the remaining
states is already sufficiently small, i.e. $S(\rho)<\log_2 d$. We
note that, as it was already mentioned in \cite{Ekert} for $d=2$,
the Isotropic twirling after each step is not really necessary and
in general increases the entropy and therefore decreases the rate
of a subsequently applied breeding/hasing protocol. However, it is
sufficient to look at the fidelity parameter $f$ in order to see
that a recurrence preprocessing improves the rates such that every
entangled Isotropic state becomes distillable.

Straightforward calculation shows that the fidelity $f'$ after
applying the recurrence method once is given by
\begin{equation}\label{fprime}
f'=\frac{1+f\big[df(d^2+d-1)-2\big]}{d^3f^2-2df+d^2+d-1},
\end{equation}
and the probability for equal measurement outcomes is
\begin{equation}\label{probrec}
p_{\text{rec}}=\frac{d^3 f^2-2df+d^2+d-1}{(d+1)^2 d(d-1)}.
\end{equation}
Note that the recurrence method alone does not lead to a non-zero
rate since in every round we destroy or discard at least half of
the resources (all the target states) and maximally entangled
states are only obtained in the limit of infinitely many rounds.
However, it holds that $f'>f$ and therefore $S(\rho')<S(\rho)$ for
every entangled isotropic $\rho$. This means that every entangled
isotropic state can be distilled with a non-zero rate by applying
sufficiently many rounds of the recurrence protocol followed by
hashing/breeding.

There are many directions in which the recurrence method can be
modified or improved. One could use more than one source state
\cite{Smolin}, a
 different bilateral operation \cite{AlberJex}, neglect the Isotropic twirling \cite{Ekert},
  apply it to other kinds of
 states \cite{AlberJex} or use target and source states with different fidelities.

\begin{figure} \begin{center}
\epsfxsize=8cm\epsffile{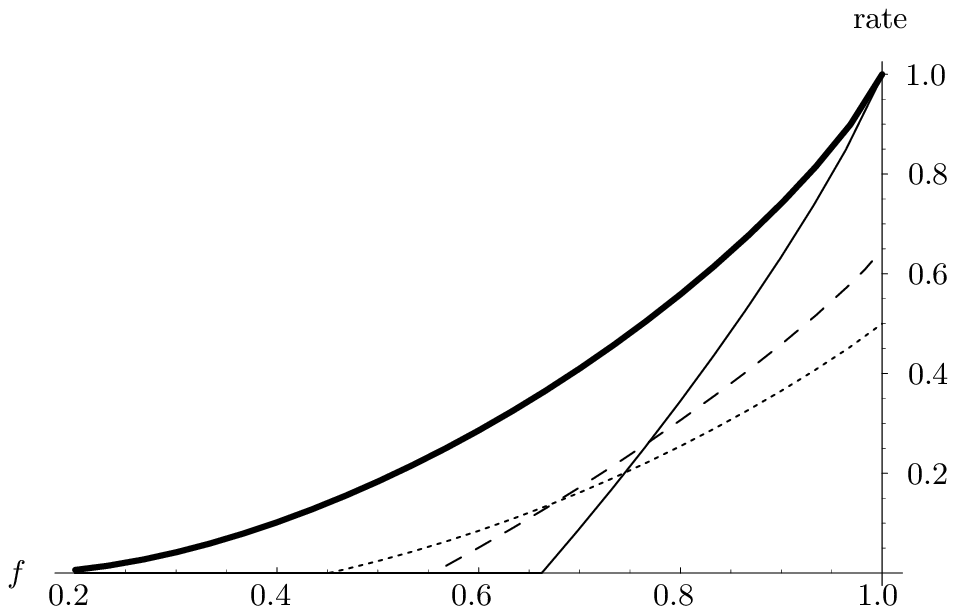}\vspace*{13pt}\caption{Normalized
rates for Isotropic states of dimension $d=7$. The upper bound
(thick) is given by the relative entropy of entanglement. The
hashing rate (solid) can be improved in the region
$S(\rho)\sim\log_2 d$ if one first applies one round of the
recurrence method (dotted) or locally projects the state onto
subspaces of dimension 3 and 4 (dashed).\label{dim5}}
\end{center}
\end{figure}

 In Fig.\ref{dim5} we have applied the method as described
 above to an isotropic state of dimension $d=7$. As expected this leads to an
 improvement of the rate in the region where $S(\rho)\sim\log_2
 d$.

\subsection{Projecting onto local subspaces}
A surprisingly efficient method for improving the distillation
rate is to project the initial state locally onto blocks of
smaller dimensions
 and to apply the hashing/breeding protocol to these subspaces.

\begin{figure} \begin{center}
\epsfxsize=8cm\epsffile{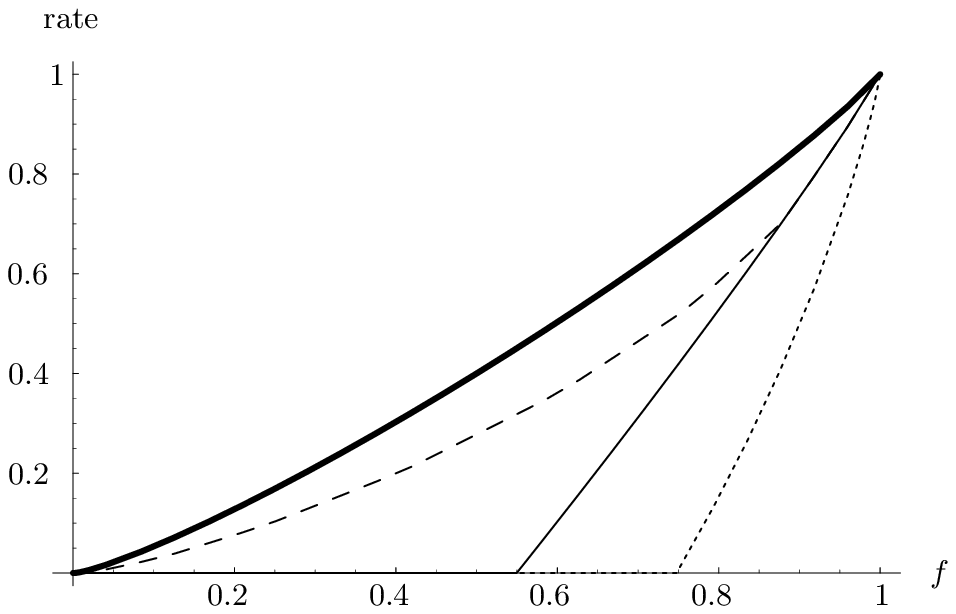}\vspace*{13pt}\caption{Normalized
rates for the case where $\rho$ consists of ten pairs of qubit
systems ($d=2^{10}$) and the overall state is isotropic. The upper
bound (thick) is given by the relative entropy of entanglement.
The hashing rate (solid) can be significantly improved when
projecting onto subspaces of dimension $2^m,\ m=1,\ldots,9$. The
dashed line is the envelope of the rates obtained in this way. The
dotted curve is the rate one would achieve when considering $\rho$
as a set of 10 pairs of qubits, twirling onto isotropic states on
$\mathbb{C}^2\otimes\mathbb{C}^2$ and applying the hashing
protocol then on the two-qubit level.\label{dim1024}}
\end{center}
\end{figure}

 Let $\{Q_i=\sum_k |k\rangle\langle k|\}$ be a set of $b$ projectors of
 dimension $q_i$
 corresponding to a local projective observable with $b$ outcomes, i.e. $\sum_{i=1}^b q_i=d$ and $ \sum_{i=1}^b Q_i={\bf 1}_d$.
 Assume further that Alice and Bob have the same observable
 $\{Q_i^A\}=\{Q_i^B\}$ and that  they apply the respective
 measurement to an isotropic state $\rho$.
 Again the state is kept if the results of the measurements coincide and discarded otherwise.
 If  both parties obtain the same measurement
 outcome $i$, which happens with a probability
\begin{equation}\label{probsub}
p_{i}=\frac{q_i}d
f+\frac{(1-f)}{(d^2-1)}\Big(q_i^2-\frac{q_i}d\Big),
\end{equation}
the state after the measurement $\rho_i=(Q_i^A\otimes
Q_i^B)\rho(Q_i^A\otimes Q_i^B)^*/p_i$ is again an isotropic state
on $\mathbb{C}^{q_i}\otimes \mathbb{C}^{q_i}$ with fidelity
\begin{equation}\label{fsub}
f_i=\Big[f\frac{q_i}d+\frac{(1-f)}{(d^2-1)}\Big(1-\frac{q_i}d\Big)\Big]/p_i.
\end{equation}
If we now apply the hashing/breeding protocol to the $b$ blocks,
the obtained rate
\begin{equation}\label{subRate}
\sum_{i=1}^b p_i \max\big[0,\log_2q_i-H(\rho_i)\big]
\end{equation}exceeds $\log_2 d-H(\rho)$ for some values of $f$ depending on the
dimensions $\{q_i\}$ of the subspaces. If the dimensions of the
blocks are about the same, then an increasing number of blocks
leads to a larger (smaller) rate for small (large) $f$.
Fig.\ref{dim5} and Fig.\ref{dim1024} show the obtained rates for
$d=7$ and the case where the two parties share 10 pairs of qubits
and the overall state is isotropic. For the latter case the rate
is already not too far below the relative entropy of entanglement.
In fact, the content of the following subsection is to sketch that
in the limit of large dimensions $d\rightarrow\infty$ the relative
distance between these quantities vanishes for all $f$.

\subsection{The limit of large dimensions}
As we have already mentioned, an upper bound for the distillable
entanglement is the {\it relative entropy of entanglement}, i.e.
the minimal distance between $\rho$ and a separable state $\sigma$
measured in terms of the relative entropy
\begin{equation}\label{RelEntdef2}
S(\rho,\sigma)=\tr{\rho\log_2\rho}-\tr{\rho\log_2\sigma}.
\end{equation}
For entangled isotropic states the nearest separable state is the
isotropic state with $f=\frac1d$ \cite{Rainsbound,VT,ESym} and the
relative entropy of entanglement is thus
\begin{equation}\label{ER}
E_R(\rho)= \log_2 d-(1-f)\log_2(d-1)-S(f,1-f),
\end{equation}
for $f>\frac1d$ and zero otherwise, where $S$ is again the Shannon
entropy. Note that in the limit of large dimensions this becomes
after normalization
\begin{equation}\label{ERlim}
\lim_{d\rightarrow\infty} \frac{E_R(\rho)}{\log_2 d}=f.
\end{equation}
The same limit, however, appears for the distillation rate
obtained with the block projection method described in the
previous subsection.

Assume that $d=2^m$, i.e. a single state $\rho$ consists of $m$
pairs of qubits, and let the local measurements have $m$ outcomes
corresponding to subspaces of equal dimensions $q_i=2^m/m$. We can
then estimate the probability of success and the achieved fidelity
(for $m\geq (1-f)^{-1}$) by
\begin{eqnarray}
p_i&\geq& \tilde{p}:=\frac{f-4^{-m}}m ,\label{ptilde}\\
f_i&\geq& \tilde{f}:=\frac{fm}{fm+1} \label{ftilde}.
\end{eqnarray}
In the limit of large $m$, the Eqs.(\ref{ptilde}, \ref{ftilde})
tell us that we gain $m$ (almost) maximally entangled states of
dimension $q=2^m/m$ with probability $f/m$. Looking at the
normalized entanglement as in Eq.(\ref{ERlim}) leads then indeed
to the same limit as for the relative entropy of entanglement.
Hence, we (relatively) approach the distillable entanglement in
the limit of large dimensions. However, the main reason for this
is, that the relative entanglement difference of two maximally
entangled states in $2^m/m$ resp. $2^m$ dimensions vanishes for
$m\rightarrow\infty$. Hence, the result is not as deep as it might
seem at first glance.

Nevertheless, projecting onto blocks of lower dimensions can
significantly improve the rates yet for finite dimensions as shown
in Figs.\ref{dim5},\ref{dim1024}.

\section{Conclusion}

Despite considerable efforts in the theory of entanglement
distillation, this field of investigation provides many open
questions even on a very basic level. Due to the complexity of the
underlying variational problem it is hard to find good upper and
lower bounds to the distillable entanglement, not to mention an
explicit calculation of this measure of ''useful entanglement``.
So far, the most important distillation protocol leading to a
non-zero rate, and thus to a non-trivial lower bound to the
distillable entanglement, has been adapted to Bell diagonal states
of two qubits. Of course, this protocol can also be applied to
higher dimensional states by either projecting down to qubits or
by  interpreting the state as a tensor product of qubits if
possible (compare Fig.\ref{dim1024}). However, both methods will
in general discard most of the entanglement.

The present article shows how to generalize the breeding and
hashing protocol to higher dimensions. The obtained rates are
optimal only for special cases, however, they provide an improved
lower bound to the distillable entanglement in general. Both
protocols consist out of two steps: first the quantum task is
translated to a classical problem and then the latter is solved.
The classical part of the protocol is already essentially optimal,
however, some information and hence entanglement is lost during
the 'translation process'.

We think that the presented results admit further generalizations
towards other classes of states and may be also with respect to
the restriction of the hashing protocols to dimensions which are
(powers of) primes.

We hope that our work initiates further investigations concerning
the distillation of entanglement, including explorations of the
implications coming from recent works in the field of quantum
error correcting codes.

\section*{Acknowledgement}
The authors would like to thank R.F. Werner , H. Barnum and D.
Schlingemann for valuable discussions. Funding by the European
Union project EQUIP (contract IST-1999-11053) and financial
support from the DFG (Bonn) is gratefully acknowledged.


\begin{references}
\bibitem{likelysequences} For a detailed description of likely
sequences we refer to \cite{Schumacher} and
\cite{BennettPRL,BennettPRA} in the context of hashing/breeding.
Since the classical part of our protocol and that of
\cite{BennettPRL,BennettPRA} is essentially the same, we refrained
from giving another detailed description of the properties of
likely sequences within the present article.
\end{references}
\end{document}